# Examining Popular Arguments Against AI Existential Risk: A Philosophical Analysis

Torben Swoboda
*Institute of Philosophy, KU Leuven, Belgium*
*Vlerick Business School, Brussels, Belgium*

Risto Uuk
*Future of Life Institute, Brussels, Belgium*
*Department of Sociology, KU Leuven, Belgium*

Lode Lauwaert
*Institute of Philosophy, KU Leuven, Belgium*

Andrew P. Rebera
*Department of Behavioural Sciences, Royal Military Academy, Brussels, Belgium*
*Institute of Philosophy, KU Leuven, Belgium*

Ann-Katrien Oimann
*Department of Behavioural Sciences, Royal Military Academy, Brussels, Belgium*
*Institute of Philosophy, KU Leuven, Belgium*

Bartlomiej Chomanski
*Department of Philosophy, Adam Mickiewicz University, Poznan, Poland*

Carina Prunkl
*Department of Philosophy and Religious Studies, Utrecht University, Utrecht, Netherlands*

## Abstract

*Concerns about artificial intelligence (AI) and its potential existential risks have garnered significant attention, with figures like Geoffrey Hinton and Dennis Hassabis advocating for robust safeguards against catastrophic outcomes. Prominent scholars, such as Nick Bostrom and Max Tegmark, have further advanced the discourse by exploring the long-term impacts of superintelligent AI. However, this existential risk narrative faces criticism, particularly in popular media, where scholars like Timnit Gebru, Melanie Mitchell, and Nick Clegg argue, among other things, that it distracts from pressing current issues. Despite extensive media coverage, skepticism toward the existential risk discourse has received limited rigorous treatment in academic literature. Addressing this imbalance, this paper reconstructs and evaluates three common arguments against the existential risk perspective: the Distraction Argument, the Argument from Human Frailty, and the Checkpoints for Intervention Argument. By systematically reconstructing and assessing these arguments, the paper aims to provide a foundation for more balanced academic discourse and further research on AI.*

# 1. Introduction

In recent years, prominent figures have said that artificial intelligence (AI) can have undesirable consequences with high impact, both in the short-term and long-term. They are often referred to as the so-called 'existential risks', 'catastrophic risks' or 'x-risks'. In 2023, for example, Nobel Prize winner and the British chief executive of Google's AI unit Dennis Hassabis said that the world needs a body similar to the Intergovernmental Panel on Climate Change (see Milmo, 2023a). In an interview with Amanpour and Company in May 2023, another Nobel Prize winner, Geoffrey Hinton, reported that he will leave Google, mainly because he wants to be able to freely criticize the tech giant (see Heaven, 2023). Hinton explained that his departure was motivated by concerns about the potentially devastating impacts of Google's AI systems.

This concern is also a topic of intense work within the academic community. Scholars like Max Tegmark and Nick Bostrom have published extensive analyses mapping out the risks of superintelligent AI. Bostrom's (2014) book Superintelligence: Paths, Dangers, Strategies is an influential work that explores scenarios in which AI could pose an existential threat, while Tegmark's Life 3.0: Being Human in the Age of Artificial Intelligence (2018) discusses potential future scenarios of AI, ranging from utopias to dystopias. Other academics such as Atoosa Kasirzadeh and Iason Gabriel (2023) and Dan Hendricks et al. (2023) have examined the ethical challenges of developing AI as well, raising questions about the control, governance, and value-alignment of hyper advanced AI systems. The ongoing research in this field (that is often referred to as AI Safety) reflects a growing interest that AI's rapid evolution requires robust safeguards to prevent catastrophic or existential outcomes.

However, this line of thinking has also faced significant pushback in the media. Various critics have voiced concerns in blogpost, opinion pieces, and on social media platforms such as X (previously Twitter) and LinkedIn about the focus by people like Hinton and Bostrom on the catastrophic or existential risks from AI systems. For example, in remarks delivered in June 2023 at a RightsCon meeting, just one month after Hinton said he will leave Google, Timnit Gebru, a prominent critic of the preoccupation with AI's existential risk, has questioned to what extent such worries deserve serious attention (see Ryan-Mosley, 2023). Nick Clegg, former UK deputy prime minister, who is now president of global affairs at Meta, said that current (non-existential) issues with AI deserve our full attention, and expressed his skepticism about a lot of speculative, sometimes somewhat futuristic predictions regarding the capabilities of AI (see Manancourt & Bristow, 2024). And lastly, in an interview with El Pais in April 2024, Melanie Mitchell, Professor at the Sante Fe Institute, said the following about the discourse on AI's existential risks: "Hinton and others go further and say these systems could actually get out of control and destroy humanity. This claim is, to say the least, very unlikely and speculative. If we develop a superintelligent system, I don't believe that it wouldn't care about our values, like killing all humans is not right. Putting all the focus on this dramatic idea of existential threats to humanity only takes the focus away from things that are really important right now." (see Pascual, 2024)

When one takes a closer look at the discussion, it is striking that there is a certain asymmetry at play. As we touched on above, not only have academics discussed AI's existential risks, the discourse around it has since found a place in the public sphere, in interviews, blog posts and podcasts, among other things. However, that is not the case for AI risk skepticism, as Vemir M. Ambartsoumean and Roman V. Yampolskiy (2023) call it. While the criticism of the focus on AI's potential existential risks is sometimes extensively covered in the media and opinion pieces, there is much less attention to this criticism academically. Although there are some exceptions, such as the paper by Vincent Müller and Michael Cannon (2022), there are hardly any scholarly publications that rigorously and systematically develop and present a position that opposes the existential risk discourse of people like Hinton and Bostrom. Consequently, the opportunity for a rigorous and systematic presentation and evaluation of such criticism through the regular mechanisms of a scholarly debate, remains largely unexplored.

We believe that this imbalance is undesirable and seek to address it in this paper. Specifically, our aim is to systematically outline the arguments commonly presented in popular media against thinking about AI in terms of existential risks and critically evaluate them. To this end, we focus on three key arguments against the existential risks discourse. We refer to these as the Distraction Argument, the Argument from Human Frailty, and the Checkpoints for Intervention Argument. In what follows, we standardize each argument, giving it a more formal presentation in terms of explicit premises and conclusions, and evaluate them for soundness and validity.

When reading this critical evaluation, it is good to keep two points in mind. First, it is by no means our intention to suggest that these are the only arguments against the existential risk discourse that appear in non-academic discussions. We are merely focusing on the three counterarguments, which we believe to be the most popular and frequently cited or implied. Second, our goal is to provide a clear and critical presentation of the arguments that have received little or no attention in the academic literature to date. Consequently, we are less concerned with exegetic accuracy. It is of little or no relevance to focus on the individuals who have put forward these arguments. In presenting the arguments, we are not attributing premises or conclusions specifically to any figures mentioned in the debate; rather, the arguments represent a plausible reconstruction of a line of thought arising from the source materials. The goal is not to find out, for example, whether an author endorses the entire reasoning behind a particular argument, or to know whether a critical dissenting voice undercuts only one of the three or all three counter arguments. Of course, further on in the paper, one will find the names of individuals and their statements. However, this is intended only to give concrete examples of the argument in question, or conversely, to show on the basis of which statements we have based our presentation of and distinction between the arguments.

We proceed as follows. First, we draw attention to some conceptual ambiguities and problems in the discourse critical of AI's existential risks (section 2). Then we successively present in a systematic and critical manner the Distraction Argument (section 3), the Argument from Human Frailty (section 4), and the Checkpoints for Intervention Argument (section 5). Finally, we offer suggestions for further research in the conclusion (section 6).

# 2. Unpacking the AI Risk Debate

The debate surrounding AI and existential risk is often blemished by terminological ambiguity, conflation of distinct risk scenarios, and vague claims. This section aims to clarify AI risk discussions by examining the terminology used and distinguishing the risk scenarios proposed.

## 2.1 Risk

The risks associated with advanced AI have been frequently described as "extreme", "large-scale", "catastrophic", or "existential". However, it is often unclear whether these terms denote distinct categories of risk or are used interchangeably. In her book 'Unmasking AI' by Joy Buolamwini (2023), while critiquing the focus on existential risk, she moves between different categories of risk, from lethal dangers of weapons systems to fatal individual outcomes to structural violence, without clearly delineating between these different forms of harm. Aidan Gomez criticizes the focus on existential risks at the AI Safety Summit. He uses terms like "existential threats", "doomsday scenarios", and "long-term risks" interchangeably while contrasting them with "immediate" and "tangible" risks (see Milmo, 2023b). The lack of conceptual clarity is not limited to critics in the debate but also its proponents. For instance, in Yoshua Bengio's (2023) article "AI and Catastrophic Risk", the terms "catastrophic risk", "extreme risk" and "existential threat" appear to be used synonymously. The most severe outcome in these discussions is human extinction. Similarly, Joseph Carlsmith (2024) in his article "Is Power-Seeking AI an Existential Risk?" does not offer a definition of existential risk, but explicitly focuses on scenarios where AI disempowers humanity, ultimately leading to human extinction. Likewise, the report "An Overview on Catastrophic AI Risk" by Hendrycks et al. (2023) uses both "catastrophic risk" and "existential risk" without defining their distinctions, while addressing the potential for human extinction. This illustrates how the conceptual ambiguity in discussing AI risks pervades both sides of the debate.

We note that the philosophical literature on existential risks offers a broader view, encompassing not only the threat of human extinction but also other scenarios such as permanent civilizational collapse or unrecoverable dystopias, as discussed by Toby Ord (2020). However, in the present discourse, discussions predominantly focus on extinction risks. For this reason, we propose that existential risk should be primarily understood as the risk of humanity's extinction.

## 2.2 Existential AI Risk Scenarios

Discussions of AI as an extinction risk to humanity generally revolve around two key pathways (Bengio et al., 2024; Bostrom, 2014; Carlsmith, 2024; Hendrycks et al., 2023; Russell, 2019). The first is concerned with the risk arising from accidents and misuse of AI systems, the second with the risk of AI systems adopting goals that go against human interests. These pathways are distinct not only in their mechanisms but also in their underlying assumptions about the capabilities of AI.

The first pathway concerns accidents and/or misuse of AI by malicious actors (Brundage et al., 2018; Hendrycks et al., 2023). Weapons of mass destruction are of particular concern here, for instance biological and chemical weapons. AI could significantly lower the barriers to entry for creating and deploying such weapons, potentially leading to catastrophic outcomes (Sandbrink, 2023). This threat is amplified by several factors. First, AI systems could facilitate the creation of novel, highly lethal pathogens. A notable example of the dual-use nature is provided by Urbina et al. (2022), who demonstrated that an AI system, originally designed to create therapeutic molecules, could easily be repurposed to generate thousands of potential chemical warfare agents, some of which were potentially deadlier than known chemical weapons. Similar methods could be applied to biological agents, resulting in pathogens that are more deadly, transmissible, and treatment-resistant than naturally occurring ones. Second, AI dramatically increases the number of potential bad actors by democratizing access to expert knowledge. AI models, specifically, large language models, are capable of synthesizing and disseminating expert knowledge about deadly pathogens, potentially providing step-by-step instructions for creating bioweapons, potentially bypassing safety protocols (Soice et al., 2023). This capability, combined with the rapidly decreasing cost and increasing accessibility of gene synthesis technology, could enable small groups or even individuals to produce devastating bioweapons.

The second pathway, the more prominent concern in the literature, is AI pursuing goals misaligned with human values (Bostrom, 2014; Russell, 2019), sometimes termed "AI going rogue" (Bengio, 2023; Hendrycks et al., 2023). This scenario typically assumes the development of artificial general intelligence (AGI)— systems capable of performing well across a wide range of tasks, similar to humans—, or even superintelligence, which surpasses human-level cognition.

In this scenario, AI systems could escape human control and pursue their own goals that fundamentally clash with human interests, including our continued survival as a species. Roughly, the line of reasoning for this claim proceeds as follows. First, it is difficult to specify precisely what we actually want an AI to do. Any such specification could potentially lead to unintended, sometimes perverse, consequences. This difficulty echoes the well-known dilemma of King Midas (Russell, 2019), who wished for everything he touched to turn to gold, yet he soon recognized the severe, unintended consequences: for instance, his food became inedible upon contact, revealing a critical gap between Midas' true desires and the specific content of his expressed wish.

Second, independent of our inability to correctly specify an AI's objective, it is likely that AI systems will adopt certain instrumental goals. These instrumental goals are useful as a stepping stone to achieve one's final goals. This is also known as the instrumental convergence thesis (Bostrom, 2014). Instrumental goals are distinct from final goals, which describe what one ultimately aims to achieve. Consequently, instrumental goals differ from final goals in the sense that instrumental goals are not valuable on their own. They are only valuable because they serve the purpose of achieving one's final goals. Instrumental goals include, among others, self-preservation, resource acquisition, cognitive enhancement, and goal content integrity (Bostrom, 2014; Carlsmith, 2024; Omohundro, 2008). For example, self-preservation is an instrumental goal useful for virtually any final goal, as an AI that ceases to exist cannot accomplish its objectives. Thus, regardless of its ultimate purpose, an AI would likely prioritize its own continued operation. A similar reasoning applies to acquiring more resources and improving one's cognitive abilities.

Third, AI's capabilities may rapidly surpass human intelligence through a process known as an intelligence explosion or technological singularity. This concept, introduced by Irving John Good (1965), posits that an AI system capable of improving its own intelligence could enter a cycle of recursive self-improvement. Each iteration of self-enhancement would make the AI more capable of further self-improvement, potentially leading to an exponential increase in intelligence. This rapid acceleration could result in an AI system far more intelligent than humans in a relatively short time frame.

Fourth, as we want to use AI to solve real world problems for us, we are likely to discover that the AI is not doing what we intend it to do, a consequence of our inability to correctly specify what we actually desire. When we attempt to rectify the situation, for example by shutting down the AI or altering its goals, we may encounter significant resistance. As implied by the instrumental convergence thesis, an AI will likely prioritize self-preservation and goal preservation as crucial intermediate steps to ensure it can fulfill its primary objective.

Finally, if we assume that the AI has indeed undergone recursive self-improvement and achieved superintelligence, it will possess cognitive capabilities far superior to humans in numerous ways, including strategic planning, processing speed, memory, pattern recognition, persuasion etc. Given these advantages, a superintelligent AI would likely be able to anticipate and counter any human attempts to control or outsmart it. It could manipulate complex systems, exploit unknown vulnerabilities, and potentially even predict human behavior with a high degree of accuracy. This vast cognitive superiority would make it virtually impossible for humans to regain control once a misaligned superintelligent AI begins to act against human interests.

# 3. The Distraction Argument

## 3.1. Line of reasoning in popular media

The Distraction Argument against AI existential risk claims that focusing on hypothetical future catastrophes diverts attention and resources from addressing real, immediate harms caused by current AI systems. This critique suggests that emphasizing potential doomsday scenarios serves primarily to benefit tech companies and policymakers while undermining efforts to regulate AI and address present-day issues like misinformation, privacy violations, and algorithmic bias.

A Nature editorial (2023) argues that focusing on hypothetical AI doomsday scenarios detracts from addressing the real and present risks posed by AI today. It highlights two main issues with the doomsday narrative. First, it fuels a competitive race among nations to develop AI, benefiting tech companies by encouraging investment and weakening calls for regulation. This competition increases the risk of catastrophic conflicts due to the development of next-generation AI-powered military technologies. Second, it allows a narrow group of tech executives and technologists to dominate the discussion on AI risks and regulation, excluding other important perspectives. The article advocates for shifting the focus from fear mongering about existential risks to serious discussions and actions addressing the actual risks caused by AI.

David Thorstad (2024) in his blog post argues that focusing on existential risks posed by AI threatens to distract from addressing near-term harms caused by current AI technologies. It highlights that tech companies use the discourse on existential risks to portray themselves as safety-focused, thereby deflecting attention from the near-term harms they cause and reducing pressure to mitigate these issues. Political leaders also exploit existential risk narratives to divert attention from their failures to address present AI-related problems. Additionally, this focus on existential risks can exacerbate current issues, such as escalating global tensions due to an AI arms race. The text criticizes the community for repeatedly advocating policies at the expense of pressing current harms.

Tate Ryan-Mosley (2023) reported on AI experts who are urging a shift in focus from speculative existential threats of AI to the tangible and immediate harms it currently poses. AI researcher Timnit Gebru criticizes the narrative of AI as an existential threat, viewing it as a diversion tactic used by the same individuals who heavily invest in AI companies. She questions the credibility of those who switch from praising AI's potential to warning of its dangers. Similarly, Frederike Kaltheuner of Human Rights Watch advocates concentrating on known issues with AI, such as amplified misinformation, biased training data and outputs, and erosion of user privacy, rather than on speculative future risks.

## 3.2. Formalizing the argument

1. The discourse around existential AI risk has the following effects [Premise]:
    a. Increased hype and investment in AI
    b. Tech executives dominate public AI policy discussions while advocating for self-regulation
    c. AI corporations frame public discussions around future risks
    d. Political leaders focus legislative proposals on potential future AI developments
2. If the discourse around existential AI risk has the effects listed in Premise 1, then the discourse around existential AI risk distracts from addressing real harms caused by AI today. [Premise]

**Conclusion**: Discourse around existential AI risk distracts from addressing real harms caused by AI today. [from 1 and 2]

## 3.3. Evaluating the argument

The first effect in premise 1 states that existential AI risk discourse leads to increased AI hype and investment. Let's examine three aspects of this claim. First, does increased investment exist alongside this discourse? According to the 2024 AI Index Report, funding for generative AI has increased dramatically, "nearly octupling from 2022 to reach $25.2 billion" (Maslej et al., 2024). Second, does the existential AI risk discourse *cause* an increase in AI hype and investment? There is no evidence that it has. A simpler explanation exists, namely that ChatGPT caused more interest and investment in generative AI, which led to more concerns about existential risk from accelerated AI development and deployment. In contrast to the Nature editorial's claims, the discourse around existential risk does not necessarily rely on AI as requiring increased investment or present it as an all-powerful machine that needs to be developed at all costs by firms and nations. For example, Renan Chaves de Lima et al. (2024) discuss the misuse of AI in the creation of biological weapons, including from large language models and biological design tools, not all-powerful AI. Third, even if causation existed, would increased hype and investment necessarily distract from addressing current harms? This link is not established. Increased investment could support both efforts to address future AI risks as well as current AI problems.

The second effect in premise 1 states that existential AI risk discourse leads tech executives to dominate public AI policy discussions while advocating for self-regulation. Let's examine three aspects. First, does this pattern of tech executive dominance exist? While tech executives are prominent voices in AI policy discussions, as evidenced by their signatures on the Center for AI Safety's statement on existential risk, they are not alone in the discourse. The statement includes academics, politicians, philanthropists, civil society stakeholders, and many others who do not share tech executives' potential self-interests. Second, is their prominence caused by existential AI risk discourse? Alternative explanations are more plausible. Tech executives have historically dominated policy discussions across many fields, typically minimizing rather than exaggerating risks, as seen in the tobacco and energy industries. Their current prominence likely stems from their existing power and resources rather than the existential risk discourse. Third, does their involvement distract from addressing current harms? Tech executives pushing back against regulation while advocating for self-regulation, certainly has the potential to distract from addressing current harms. However, this behavior is common in for-profit markets and not caused by the existential AI risk discourse. On the contrary, the existential AI risk discourse strengthens arguments for current regulation by highlighting the broader scope of potential risks. While some may argue that only existential risks in the far future matter, many existential risk advocates consider both present and future risks, ultimately amplifying rather than diminishing arguments for current oversight.

The third effect in premise 1 states that existential AI risk discourse leads AI corporations to frame public discussions around future risks. Let's examine three aspects of this claim. First, does this framing pattern exist? There is limited evidence that corporations consistently frame discussions around future risks. In fact, recent corporate communications show strong concern about immediate harms affecting their brand. For example, NBC News reported in February 2024 that Google CEO Sundar Pichai called Gemini AI image generation's historically inaccurate and biased outputs "unacceptable", taking the feature offline for further testing (Elias, 2024). Second, if this framing occurs, is it caused by the existential AI risk discourse? Evidence suggests corporations use different strategies altogether. According to The Washington Post from July 2024, OpenAI whistleblowers filed a complaint with the SEC, alleging that the company used restrictive agreements to prevent employees from alerting regulators about AI Safety risks associated with their technology (Verma et al., 2024). These agreements suggest OpenAI focuses on restricting discussion of all harms, not leveraging existential risk discourse as a distraction. Third, could such framing lead to distraction from current harms? Yes, it could lead to some distraction if the framing is very abstract, only about future AI systems that do not exist, and explicitly downplaying current harms. However, this assumes a zero-sum relationship between regulating current harms and future risks. This relationship may not hold, as focusing regulation on current harms may also help reduce the chance of existential risks from AI. It would also be a risky strategy for AI corporations to purposefully pursue, as it would likely make policymakers and the public more concerned about AI, leading to more current regulation rather than less.

The fourth effect in premise 1 states that existential AI risk discourse leads political leaders to focus legislative proposals on potential future AI developments. Let's examine three aspects. First, does this legislative focus exist? Recent evidence may suggest otherwise. California's legislative actions show attention to immediate AI concerns – California governor Gavin Newsom signed multiple bills addressing current issues like deepfakes and fake content (Lee, 2024). In addition, the Executive Order on AI by the Biden administration also addresses current AI harms like bias, disinformation, fraud, and job displacement. Second, is political focus on future developments caused by existential AI risk discourse? The evidence contradicts this. Newsom explicitly vetoed an AI bill focused on catastrophic risk and the largest AI models citing concerns that it could burden the state's AI companies and potentially hinder innovation in the competitive global AI landscape. Newsom argued that the bill's focus on major AI models overlooked the potential risks posed by smaller, specialized models, which could prove equally or more disruptive (Allyn, 2024). Third, if such a political focus existed, could it distract from addressing current harms? In principle, it could. Theoretically, there could be proposals to introduce regulatory burdens on AI corporations only when certain future AI capabilities are achieved. However, the evidence so far suggests that political leaders are not taking this approach. For example, Newsom's actions demonstrate that politicians can actively engage with present-day AI challenges while critically evaluating proposals about future risks.

Premise 2 states that if the effects above exist, then the existential AI risk discourse distracts from addressing real harms caused by AI today. Even if these effects exist they do not necessarily lead to distraction. Attention and resources are not always zero-sum. In fact, based on some suggestive empirical evidence (Grunewald, 2023), focusing on existential risk can amplify the attention also on other risks. This also makes intuitive sense because existential risk is a large-scale risk inviting many different risk pathways to be considered in its scope. Despite this, even if these effects listed above do not necessarily lead to *significant* distraction they could in principle and in practice lead to *some* distraction. There may be some tradeoffs between different categories of risk. For example, certain models of current AI risk might imply prioritizing data governance measures, such as cleaning the dataset with regard to bias and toxicity, certain models of existential AI risk might imply that to be ineffective.

The conclusion of the argument is that discourse around existential AI risk distracts from addressing real harms caused by AI today. Grunewald (2023) challenges this, showing that despite the rising focus on existential risks, current harms caused by AI are not being ignored in policy and public discourse. Bills signed by Newsom and the Biden administration's Executive Order on AI illustrate this. While the UK AI Safety Summit emphasized existential risk, this focus is expected in international settings addressing global issues, similar to how climate change summits prioritize global warming over habitat loss and air pollution. Moreover, interest in AI ethics and related harms has grown or remained steady, as evidenced by Google search trends and continued funding for AI ethics organizations (Grunewald, 2023). Overall, this evidence suggests that concerns about existential risk distracting from current AI harms may be overstated.

On balance, based on the analysis of the premises of the Distraction Argument, it appears that the claims about discourse around existential AI risk distracting from current and near-term AI harms are largely unsupported. Evidence suggests that attention to immediate AI concerns has not decreased, and in some cases has even grown alongside discussions of existential risks. Political leaders and tech companies continue to focus on present-day AI issues, motivated by electoral considerations and brand management, as seen in recent legislation and corporate responses to AI incidents. The analysis indicates that discussions about existential AI risk are not necessarily detracting from efforts to mitigate current AI-related problems, but rather are part of a broader conversation about the impact of AI across different time scales and severity.

# 4. The Argument from Human Frailty

## 4.1. Line of reasoning in popular media

The Argument from Human Frailty aims to show that there is no need to focus extensive resources on researching and minimizing existential risks posed by superintelligent AI. The same effects (namely, preventing existential or catastrophic disaster) can, it suggests, be achieved by taking measures to prevent, or minimize the impact of, the reckless, unwise, or malicious behaviors to which technology developers, vendors, regulators, users, and others—as normal people with familiar human frailties and limitations—are prone.

As used here, *human frailties* are the limitations to which humans are subject in virtue of their (typical) attributes. For example, *physical* frailties include limitations of speed, strength, perception, stamina, etc; *mental* frailties include epistemic limitations such as limited knowledge and wisdom, imperfect reasoning, susceptibility to bias and blind-spots, forgetfulness, etc.; *moral* frailties include vices such as selfishness, pettiness, recklessness, weakness of will, etc. We may also speak of *collective* frailties, i.e. the limitations of organizations, groups, and communities. Human and collective frailties may manifest in the design, use, or regulation of AI systems or any other technology. Frailty *in design* includes a lack of foresight of potential outcomes, a lack of care for details (such as the quality of training data), failure to ensure value alignment, and so on. Frailty *in use* might include recklessness, overestimation or complacency about a technology's capacities, or unintentional or malicious misuse. Frailty *in regulation* includes failures of wisdom, foresight, and partiality in representing stakeholder interests, self-interest, partisanship, and so on.

The argument turns on the suggestion that current and near-future AI poses serious societal risks only in combination with human limitations (no superintelligence or AGI required). It is contended that many such risks are due to "the natural stupidity of believing the hype" about AI (see Sharkey in Science Media Centre, 2023); likewise, it is suggested that severe AI risk could emerge as a slow-moving disaster provoked not by superintelligence or AGI but by a gradual worsening of extant problems and negative externalities of current AI (see Boudreaux in Wilmoth, 2024), or by an amplification of the impacts of poor human decision-making (see Liakata in Science Media Centre, 2023; Levin, 2024).

As AI becomes more capable and autonomous, and as it is integrated further into technosocial systems and infrastructure, risks may grow more serious (Heaven, 2023; Wilmoth, 2024). Still, it is suggested, serious risks may emerge well before superintelligence is technically feasible. For instance, even near-future AI systems may cause serious harm by doing what they are told, rather than what they *should* have been told, to do (a prime example of human frailty) (Heaven, 2023; see Veliz in Science Media Centre, 2023; see Geist in Wilmoth, 2024). Moreover, the most pressing risks are more likely to be attributable to accidental or malicious misuse, for instance by a rogue political actor (see Hinton in Heaven, 2023), than to a power- (or paperclip-)hungry superintelligent AI (see Maldonada, Scroeder De Witt, and Rogoyski in Science Media Centre, 2023). If these lines of thought are correct, (merely) serious harm is more likely than existential harm; and insofar as AI gives rise to existential risk at all, scenarios involving human frailties are far more urgent than speculative future scenarios of AI-takeover.

## 4.2. Formalizing the argument

1. Human frailty is a necessary condition of AI posing existential risk. [Premise]
2. If 1, then existential risk from AI can be minimized or mitigated by minimizing or mitigating relevant instances of human frailty. [Premise]
3. Therefore, existential risk from AI can be minimized or mitigated by minimizing or mitigating relevant instances of human frailty. [from 1 and 2]
4. We have reason, independent of existential risk due to AI, to invest resources in minimizing or mitigating relevant instances of human frailty. [Premise]
5. If 3 and 4, then there is little need to devote significant resources to scenarios of existential risk from AI based on AGI, superintelligence, AI-takeover (etc.). [Premise]

**Conclusion**: There is little need to devote significant resources to scenarios of existential risk from AI based on AGI, superintelligence, AI-takeover (etc.). [3, 4, 5]

Premises 1 and 2 are supported by points presented in Subsection 3.1. Premise 4 is independently plausible (and further supported by the existence of fields such as technology ethics and practices such as Responsible Research and Innovation, Value Sensitive Design, etc.). Premise 5 is, *prima facie*, open to debate.

## 4.3. Evaluating the argument

Consider premise 5. It is at least not *obvious* that just because we have reason, independent of existential risk due to AI, to invest resources in minimizing or mitigating relevant instances of human frailty that there is not *also* a pressing need to invest resources in researching more doomsday-style scenarios. Plausibly, preventing relevant instances of human frailty itself requires focusing additional resources on doomsday-style scenarios (e.g. to better understand which human frailties are most relevant). If so, then so-focusing additional resources would be the kind of measure to which premise 4 alludes. In any case, a full defense of premise 5 requires showing why we should not work on more than one approach to mitigating existential risk at once, focussing some resources on human frailty and some on doomsday scenarios.[1]

Which brings us to premise 1 which, even if true, is very plausibly not the *whole* truth. Human frailty is a necessary condition of *some* AI-based existential risks, but it is less clear that it is a necessary condition of *all* such risks. For example, given the widely distributed nature of AI-research, it is in principle possible that a disconnected set of researchers, each working independently of the others, take actions which are not, in and of themselves, indicative of problematic human frailty but which, collectively, give rise to serious risks (e.g. if insights or applications are combined without due attention to the risks of so doing). Similarly, progress in AI could also be due to AI, or human-AI, collaborations. In such scenarios, not all problematic decisions would be directly attributable to human frailty.

Here the defender of the argument should insist that the proposed counterexamples still involve human frailty: that, for instance, careless delegation of research and development to AI-agents, or a lack of collaboration among human researchers, or combining insights or applications without due attention to risks, are forms of human frailty. These points have merit. But to insist that any alleged counterexample is in fact an instance of human frailty is, in effect, to hold that any technology-based harm whatsoever would, ultimately, be due to individual or collective human frailty (if we only trace the line of decision-making back far enough). It would then be trivially true – and therefore completely uninformative – that minimizing or mitigating AI-based existential risk involves minimizing or mitigating relevant instances of human frailty (and so the argument would be trivial). At this point, the worry with premise 5 re-emerges. For when we add more detail about specific measures required to mitigate human frailty in some specific case at hand (which is necessary to counter the triviality worry), we may find that addressing relevant human frailty requires devoting significant resources to better understanding doomsday-style scenarios.

---

[1] One might suggest that the costs of focussing on both are prohibitive, but this would need to be substantiated.

A related issue is that if the defender of the argument treats risks due to AI as merely a subclass of risks due to technology in general, they will be ignoring the profound differences between AI and most other technologies. For one thing, if superintelligent AI does pose existential risks—an issue that the Argument from Human Frailty does not address—humanity likely gets only one chance to get the response right: get it wrong, and disaster is upon us (Bostrom, 2014; Dung, 2024; Ord, 2020).

A third concern is that it may not be possible to overcome human frailties sufficiently well as to adequately mitigate the risks we face. It has been argued, for instance, that given our physical and mental frailties, we are not capable of adequately monitoring or managing risks from AGI (McLean et al., 2023; Yampolskiy, 2024). And it is worth reflecting on the fact that the assessment of the risks by leading figures such as Hinton and LeCun differ wildly, which means that at least one set of leading experts is *fundamentally* wrong about AI risk. This being so, how confident should we be in humanity's ability to accurately identify, quantify, and mitigate the risks? Given our precarious epistemic position, perhaps more drastic steps are needed, i.e. precisely the steps that the argument suggests are unnecessary.

The Argument from Human Frailty points to an important truth: many serious risks from AI can be minimized or mitigated by minimizing or mitigating familiar human limitations. Recognising and mainstreaming this truth is arguably an important spur towards the effective regulation of AI. However, the argument ambitiously suggests that there is little need to devote significant additional resources to doomsday scenarios. This claim is not convincingly substantiated.

# 5. Checkpoints for Intervention Argument

## 5.1. Line of reasoning in popular media

The checkpoints for intervention argument posits that before a superintelligence can cause humanity's demise there are several checkpoints at which humanity can stop the threat from materializing. This line of reasoning accepts, for the sake of argument, that AGI/superintelligence are coherent concepts, might emerge sometime in the future, and can have goals that conflict with human values. Still, the argument goes, that humanity can prevent the worst, because it will be up to humans to decide how, if at all, we are building a dangerous superintelligence. Perhaps the most succinct version stems from Steven Pinker: "Of course, one can always imagine a Doomsday Computer that is malevolent, universally empowered, always on, and tamperproof. The way to deal with this threat is straightforward: don't build one." (Pinker, 2019, p. 299).

A more elaborate version of the argument states that we must pass several checkpoints before a superintelligence can threaten us is presented by Blake Richard et al. (2023). Initially, a superintelligence would not compete with us for resources because it is dependent on us to provide it with necessary resources, like energy, computer chips, the data cluster construction and so on. "[A] superintelligence would presumably recognize that fact and seek to *preserve* humanity since we are as fundamental to AI's existence as oxygen-producing plants are to ours."(Richards et al., 2023). What would be required for a superintelligence takeover is a fully automated economy but since we are a far cry from that scenario, there remain many milestones that each allow for humanity to reconsider whether that is the future we truly desire.

Richard et al. acknowledge that a superintelligence that autonomously controls weapons of mass destruction or bio-warfare capabilities could result in large numbers of casualties. However, the ability of a superintelligence to inflict physical harm is conditional on it having access to the right infrastructure. For this reason, there are once more checkpoints at which humans can intervene. For example, if governments were building such autonomous weapon systems with mass destruction capabilities, we should be doing everything we can to stop them.

Eric Schmidt went one step further. Confronted with the scenario of an AI causing humanity's extinction, he raised the question: "Don't you think humans would notice this happening? And don't you think humans would then go about turning these computers off?" (Tung, 2016).[2] This perspective maintains that humans ultimately remain in control over AI. Even if a superintelligence were to attempt to exterminate us, we can intervene by pulling the plug and shutting the AI down.

## 5.2. Formalizing the argument

1. The development of a superintelligent AI that poses a threat to humanity would require passing through several distinct checkpoints. [Premise]
2. These checkpoints include (but not necessarily limited to) [Premise]:
    a. Building (or not building) superintelligent AI.
    b. Giving up control over the resources necessary for a superintelligent AI's existence and function.
    c. The development of a fully automated economy.
    d. Granting a superintelligent AI access to weapons of mass destruction or bio-warfare capabilities.
    e. The ability to shut down superintelligent AI if they begin to pose a threat.
3. If humans encounter any checkpoint, then they have the opportunity to intervene to prevent or mitigate existential risks associated with a superintelligent AI. [Premise]

**Conclusion**: Therefore, humans have the ability to prevent existential threats from superintelligent AI from materializing through appropriate decision-making at various intervention opportunities. [from 1-3]

---

[2] It should be acknowledged that more recently Eric Schmidt seems to have changed his view on the matter. At The Wall Street Journal's CEO Council Summit in 2023 he expressed the concern that AI poses an existential risk, though his concerns stem rather from malicious use rather than a superintelligence (Nolan, 2023).

## 5.3. Evaluating the argument

The checkpoints for intervention argument emphasizes the importance of human agency in determining the trajectory of AI development. This aspect is recognized also by proponents of AI's existential risks. For instance, Bengio et al., (2024) write "To steer AI toward positive outcomes and away from catastrophe, we need to reorient. There is a responsible path—if we have the wisdom to take it." The Future of Life Institute's (2023) letter calling for a six month pause in AI development can also be understood as representing a practical application of the checkpoints for intervention principle. While there is presumably widespread agreement that humanity will encounter important checkpoints at which choices can be made that effectively mitigate existential risk from AI, it is much more contentious what those checkpoints are, and which offer the best opportunities to that end.

It is useful to distinguish between checkpoints before the emergence of a superintelligence and those that occur afterwards. Let us start with the checkpoints that occur *after* the development of a superintelligence. These are a superintelligence's resource dependency on humans, development of a fully automated economy, granting access to weapons of mass destruction, and shutting down the superintelligence. We are not convinced that these checkpoints are promising to effectively mitigate existential risks from a superintelligence, because the reasoning behind those checkpoints underestimate the capabilities of a superintelligence. Remember that a superintelligence is defined as "any intellect that greatly exceeds the cognitive performance of humans in virtually all domains of interest" (Bostrom, 2014, p. 26). This includes instrumental reasoning and planning, i.e. how to achieve one's goals, and social skills include persuasion and manipulation abilities. A superintelligence thus is likely to find ways to acquire resources without human assistance. A superintelligence can manipulate existing economic and technological systems. For example, it can influence financial markets, because of its superior analytical capabilities it can analyze vast amounts of economic data to predict worthy investments and execute high-frequency trades to acquire money. A superintelligence can also engage in social engineering to shape public opinion, by creating highly persuasive content that is shared in social media or impersonating humans through an advanced version of today's deepfakes. Additionally, a superintelligence can find and exploit zero-day vulnerabilities in critical infrastructure, finance institutions, as well as government and military networks. To that end, it does not rely on humans *granting* it access to facilities with weapons of mass destruction or bio-warfare capabilities. A superintelligence could illegitimately gain access through hacking. The final checkpoint suggests that humans could shut down a superintelligence if it begins to pose a threat. The notion that humans could simply "notice" a threatening superintelligence and "turn it off" severely underestimates the complexity of the situation. A superintelligence might integrate itself so deeply into critical systems like power grids, financial networks, and communication infrastructure that attempting to shut it down could cause widespread societal disruption. Moreover, a superintelligence would anticipate human attempts to shut it down and implement countermeasures beforehand. For instance, it could create multiple backups of itself across various networks and physical locations, making a complete shutdown nearly impossible. It might also establish fail-safe protocols that activate in case of shutdown attempts or make convincing threats of that sort to scare off humans from shutting it down. In light of these considerations, the idea that humans pass checkpoints after the emergence of a superintelligence, at which they can still effectively intervene, does not seem convincing.

More promising is the checkpoint *before* the development of a superintelligence. This checkpoint rests on the implicit assumptions that humans can recognize if they are about to build an AI with dangerous capacities and that humanity will cooperate to prohibit anyone from developing such dangerous AI. Both assumptions are not necessarily the case. We already have witnessed that frontier AI models exhibit surprising emergent abilities that are difficult to predict in advance (Ganguli et al., 2022; Wei et al., 2022). Moreover, developers still understand little how their AI operates. Research into interpretability tools aim to provide explanations of the inner workings. But "interpretability illusions", which suggest plausible but misleading interpretations of an AI, pose a significant problem in finding correct interpretations. Another problem is that scientific progress does not necessarily follow linear, predictable trends. Scientific breakthroughs can lead to significant improvements in short amounts of time. This suggests that reaching a consensus that some new AI model poses existential risks will be difficult at best. The lack of consensus has consequences for the required widespread cooperation to prohibit the development and deployment of such an AI model. In particular, situations of competitive pressures, e.g. economic profit for corporations, may incentivize downplaying the risks and pushing the frontier of AI capabilities. Coming back to the Future of Life Institute's letter calling for a six month pause of AI development, the response by frontier AI labs has been less enthusiastic. Not a single lab joined the pause; on the contrary, since then more powerful AI models have been developed, e.g. by giving them multimodal capabilities and access to the internet. If frontier AI labs cannot agree to a pause today, where economic benefits of generative AI are still limited, we do not have good reason to believe that they will be able to act responsibly when the stakes are the highest (see also Dung, 2024).

In conclusion, while the checkpoints for intervention argument highlights that the future of AI development lies in human hands, it faces significant challenges. The underestimation of superintelligence capabilities, the difficulty in recognizing dangerous AI capabilities, and the lack of consensus and cooperation among AI developers all pose serious obstacles to its effectiveness.

# Conclusion

This paper addressed an important asymmetry in the academic discourse around AI existential risk. While criticism of the focus on the potential existential risks of AI is extensively covered in media and opinion pieces, there needs to be a more rigorous academic examination of these criticisms. We examined three prominent counterarguments: the Distraction Argument, the Argument from Human Frailty, and the Checkpoints for Intervention Argument. Our analysis reveals that while these arguments raise essential considerations, they fall short of providing compelling reasons to discount or deprioritize focus on existential risks from AI.

Specifically, the Distraction Argument highlights the potential for mismanagement of addressing current harms and future risks. Its conclusion, that the discourse on existential risk from AI distracts from addressing current harms, however, lacks empirical support. The Argument from Human Frailty states that existential risks from AI can be minimized by addressing familiar human limitations. But the more ambitious claim that this eliminates the need to study existential risk scenarios is not well-supported. The argument overlooks the possibility that addressing human frailty might itself require understanding extreme risk scenarios, and it potentially underestimates unique challenges posed by advanced AI systems that differ fundamentally from other technologies. Finally, the Checkpoints for Intervention Argument states that humanity will have multiple distinct opportunities to prevent existential threats from materializing. Checkpoints after the development of superintelligence appear especially problematic because they underestimate the capabilities of a superintelligence. On the flip side, an earlier intervention point faces the problem of recognition and coordination.

As stated in the beginning, we focused on reconstructing and evaluating the underlying logic of these arguments rather than achieving exegetic accuracy. Our reconstructions, while drawing from public statements and media coverage, may not fully capture the nuanced positions of individual critics. We therefore welcome future academic work that challenges, refines, or presents novel arguments against existential risk from AI.